\documentclass[aps,superscriptaddress,showpacs,amsmath,amssymb,amsfonts,altaffillsymbol,twocolumn,notitlepage,prl]{revtex4-1}

\usepackage{graphicx}
\usepackage{float}
\usepackage{dcolumn}
\usepackage{bm}
\usepackage{color}

\graphicspath{{./figures/}}

\begin{document}

\title{Metastability as a mechanism for yielding in amorphous solids under cyclic shear}

\author{Muhittin Mungan
}
\email[Corresponding author: ]{mungan@iam.uni-bonn.de}
\affiliation{Institut f\"{u}r angewandte Mathematik, Universit\"{a}t Bonn, Endenicher Allee 60, 53115 Bonn, Germany}

\author{Srikanth Sastry
}
\email[Corresponding author: ]{sastry@jncasr.ac.in}
\affiliation{Jawaharlal Nehru Centre for Advanced Scientific Research, Jakkar Campus, 560064 Bengaluru, India}

\begin{abstract} 
We consider the yielding behavior of amorphous solids under cyclic shear deformation and show that it can be mapped into a random walk in a confining potential with an absorbing boundary. The resulting dynamics is governed by the first passage time into the absorbing state and suffices to capture the essential qualitative features recently observed in atomistic simulations of amorphous solids. Our results provide insight into the mechanism underlying yielding and its robustness. When the possibility of activated escape from absorbing states is added, it leads to a unique determination of a threshold energy and yield strain, suggesting thereby an appealing approach to understanding fatigue failure.
\end{abstract}
\maketitle

Yielding in amorphous solids is of importance for understanding their behaviour under applied stress in a variety of materials science and soft matter contexts  \cite{Schuh2007,Bonn2017c}, and has been investigated actively in recent years through a variety of theoretical approaches and computer simulations \cite{Falk2011,Dasgupta2012,JinWyart2016,regev2015reversibility,Itamar2016yielding,Jin,parisi2017shear,Urbani2017b,leishangthem2017yielding,parmar2019strain,Ozawa2018a,bhaumik2019role,Radhakrishnan2016b,Popovic2018a,Nicolas2018,Fielding2020,ParmarReview2020}. Simulations have been performed most often employing the athermal quasistatic (AQS) shear  \cite{shi2007,leishangthem2017yielding,Ozawa2018a,parmar2019strain,bhaumik2019role,Yeh2019}, although not exclusively \cite{PRIEZJEV2018,parmar2019strain,VishwasPRE2020}. Several studies have focused on yielding behaviour of model glasses under cyclic shear deformation \cite{Fiocco2013,regev2013,PRIEZJEV2013,regev2015reversibility,leishangthem2017yielding,parmar2019strain,bhaumik2019role,Yeh2019}, and indicate that yielding occurs as a sharp, discontinuous transition. As in the case of uniform shear \cite{Ozawa2018a,Popovic2018a,Fielding2020}, the yielding behavior displays a strong dependence on the degree of annealing \cite{bhaumik2019role,Yeh2019} and has led to the following ``phase-diagram": 
With repeated cycles of strain, the energies, as well as  other properties, of glasses change and reach a steady state. As a function of applied strain amplitude, the steady state energies of initially poorly annealed glasses  first decrease towards a common threshold energy, and then 
increases discontinuously 
upon yield. The number of cycles to reach steady states increases as the yielding transition is approached.  For higher degrees of annealing, {\em i.e.} lower initial energies, the energies from cycle to cycle do not change until yielding where an abrupt transition to the steady-state occurs. This transition becomes more discontinuous for larger annealing. The properties of the yielded glasses do not depend on the initial degree of annealing, and display strain localisation \cite{parmar2019strain}. 

Simplified models describing the observed yielding behavior under cyclic shear, particularly athermal quasistatic shear, have recently been investigated \cite{MunganWitten2019,regevLemberg2020,sastry2020mesoland,liu2020oscillatory,Maloney2021}. In \cite{sastry2020mesoland}, one of us considered the behaviour 
of a family of {\it mesostate} models, meant to describe shear induced changes of state within a single mesoscale block. Remarkably, key features of yielding under cyclic AQS shear and its dependence on the degree of annealing observed in simulations are reproduced robustly by these models.

Two observations in \cite{sastry2020mesoland} motivate the present study. (1) Starting with a state of some initial energy and considering the outcome of applying a single cycle of shear there are two possibilities. If the energy at the end the cycle is lower than a limit value that is set by the amplitude of shear, the system is stable with respect to further cycles of shear. Otherwise, subsequent cycles of shear will induce further transitions. In the latter case it was observed that the distribution of energies reached at the end of a cycle does not depend significantly on their values at the beginning of the cycle. (2) Assuming that this distribution is the invariant distribution emerging under a stochastic dynamics, below the yielding point, the average time to reach a stable final state could be accurately predicted by evaluating the time required to reach an absorbing boundary.

These observations suggest that it is useful to model the evolution of the state of an amorphous solid, from cycle to cycle, as a stochastic process that is governed by an invariant distribution, and in the presence of an absorbing region whose extent is determined by the applied shear. We present such a minimal model, that reproduces key qualitative features of the phase diagram of a sheared amorphous solid under cyclic shear. As we discuss, the observed behaviour arises as a manifestation of metastability \cite{AntonBook}: The system resides in a steady-state determined by the invariant distribution, with a low but finite transition probability into an absorbing state which is tuned by the imposed shear.

The advantage of studying minimal models is two-fold: (i) It permits a rigorous evaluation of emergent properties of the cyclic shear process and provides insights into the origin of their robust features. (ii) It also permits the inclusion of other features of dynamics in a systematic way. We demonstrate this by considering the role of activated processes that may destabilize the absorbing state, and argue that this offers an appealing approach to understand fatigue failure, a phenomenon of great practical importance, wherein a solid may fail after a large but finite number of cycles of deformation well below the yielding point. We make these considerations more precise below. \\

\noindent{\it Mesostate Model of a Sheared Amorphous Solid:}
To motivate our approach, we consider the {\em regular model} that was introduced in \cite{sastry2020mesoland}.
The set of possible states for the amorphous solid is labeled by $(\epsilon,n)$, where $\epsilon \ge 0$ is an energy and $ n = 0, \pm 1, \pm 2, \ldots $ label distinct states of energy $\epsilon$. Each pair $(\epsilon,n)$ describes
the net energy of the system when subjected to a shear strain $\gamma$ given by 
\begin{equation}
 E_{\epsilon, n}(\gamma) = - \epsilon + \frac{\kappa}{2} \left  ( \gamma - \gamma_{\epsilon,n} \right )^2, \quad 
 \gamma^-_{\epsilon,n} < \gamma < \gamma^+_{\epsilon,n}, 
 \label{eqn:mesoEnergy}
\end{equation}
where $\kappa$ is a constant, $\gamma_{\epsilon,n} = 2n \sqrt{\epsilon}$ the strain at which the energy is minimum, and $\gamma^\pm_{\epsilon,n} = \gamma_{\epsilon,n} \pm \sqrt{\epsilon}$ mark the range in strain over which the system in state $(\epsilon,n)$ responds purely elastically. 
In \cite{mungan2019networks} we have called the elastic branches $(\epsilon,n)$  {\em mesostates}, and we therefore refer to this model as a mesostate model. Under athermal oscillatory strain, the energy of the system in mesostate $(\epsilon,n)$ will vary according to Eq. \eqref{eqn:mesoEnergy}, as long as the applied shear remains within the stability interval $(\gamma^-_{\epsilon,n}, \gamma^+_{\epsilon,n})$. When the boundary $\gamma^{\pm}_{\epsilon,n}$ is reached, a (plastic) transition to a new mesostate $(\epsilon', n')$ has to occur. For athermal dynamics, energy is only dissipated and thus  $ E_{\epsilon', n'}(\gamma) <  E_{\epsilon, n}(\gamma)$. However, this condition does not exclude  transitions where 
$\epsilon' < \epsilon$ i.e. to higher energies of well minima in Eq. \eqref{eqn:mesoEnergy},
and in fact such transitions are  {\it essential} for non-trivial dynamics \cite{sastry2020mesoland}. 
Considering the mesostates attained at the end of each cycle, the dynamics is effectively one dimensional (since states stable at zero strain have $n = 0$), and continues until a mesostate $(\epsilon_f,n_f=0)$ is reached such that $\gamma^2 < \epsilon_f$, and the system will respond purely elastically to subsequent cycles of shear and no further transitions occur. We shall call this the {\em absoprtion condition}.  The particular sequence of transitions depends on the details of the dynamics of the mesostate model. As motivated earlier, however, we assume that it is governed by a stationary stochastic process, along the energy axis $\epsilon$. 
We introduce next an idealized model whose stochastic dynamics describes a random  walk along the energy axis that is trapped in a confining region whose boundary marks the absorbing regime.

\noindent{\it The Ehrenfest model:} 
We consider a reversible nearest-neighbour Markov chain, having a finite state space $\mathcal{E}$ of $2\mathcal{N} + 1$ mesostates, with energies $\epsilon_k = \frac{k}{2\mathcal{N}}$, $\quad k = 0, 1, 2, \ldots 2\mathcal{N}$. The one-step transition probabilities of the Markov chain are assumed to be independent of $\gamma$, $P(X_{t + 1} = \epsilon_j \vert X_{t} = \epsilon_k ) = p(\epsilon_k, \epsilon_j) \equiv p_{k,j}$ with 
\begin{equation} 
 p_{k,j} = \left(1 - \frac{k}{2\mathcal{N}} \right ) \, \delta_{k+1,j} + 
 \frac{k}{2\mathcal{N}} \, \delta_{k-1,j}.
 \label{eqn:EhrenfestTrap}
\end{equation}
Taking $k$ to represent the number of balls (out of  $2\mathcal{N}$) in one of two urns, the Markov chain generated by Eq.~\eqref{eqn:EhrenfestTrap} is a description of the Ehrenfest Urn model \cite{kac1947random,bellman1951recurrence}: at each time step we pick one of the $2\mathcal{N}$ balls uniformly and at random and transfer it to the other urn. The Markov chain has invariant measure $\mu(\epsilon_k)$, given by the binomial distribution
\begin{equation}
 \mu_k = \mu( \epsilon_k) = \left ( 
 \begin{array}{c}
  2\mathcal{N} \\ k
 \end{array}
 \right ) \, \frac{1}{2^{2\mathcal{N}}},
\end{equation}
as is readily checked. $\mu_k$ is unimodal with mean  $\epsilon_{\rm ss} = 1/2$, the {\em steady-state energy} (see below), and typical  fluctuations around $\epsilon_{\rm ss}$ 
being of order $1/\sqrt{\mathcal{N}}$.  Given $\gamma$, the absorption condition is obeyed by  the set of states  $A_\gamma = \{ \epsilon_j :  \gamma^2 < \epsilon_{j}\}$. For the nearest-neighbour random walk initialized outside $A_\gamma$, the absorbing condition is satisfied when the mesostate in $A_\gamma$ with  smallest energy is reached. We denote this mesostate as  $\epsilon_\gamma = k_\gamma /(2\mathcal{N}) \approx \gamma^2$, for $\mathcal{N}$ large. 

\begin{figure}[t!]
\begin{center}
   \includegraphics[width=\columnwidth]{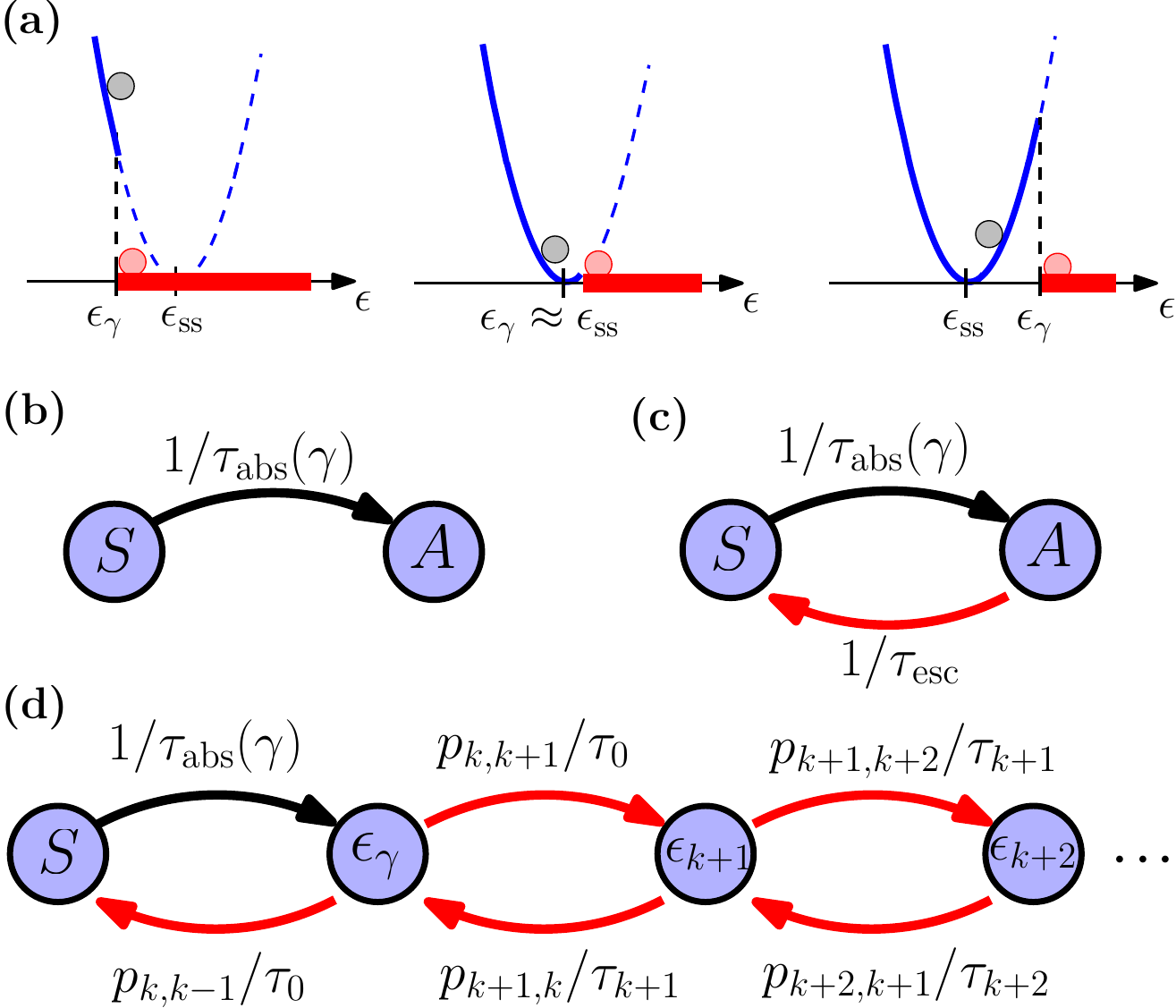} \\
    \vspace{2ex}
    \caption{(a) The dynamics of the mesostate model under cyclic shear at strain amplitude $\gamma$ can be viewed as the motion of the state point or a particle (gray circle) diffusing along the energy axis $\epsilon$. The particle is confined by an effective potential (blue) with minimum at $\epsilon_{\rm ss}$ and a $\gamma$-dependent absorbing region $\epsilon \ge \epsilon_\gamma$ (red region). The dynamics depends on whether (from left to right) $\epsilon_{\gamma} < \epsilon_{\rm ss}$, $\epsilon_{\gamma}  \approx \epsilon_{\rm ss}$, or $\epsilon_\gamma  > \epsilon_{\rm ss}$, the case of metastability where the particle is trapped near $\epsilon_{\rm ss}$ and the transition into the absorbing region is a rare event. Dividing the states in to steady state (S) and absorbing (A) regions, (b) illustrates the athermal cases with transitions from $S$ to $A$ with rate $1/\tau_{\rm abs}(\gamma)$, (c) illustrates the inclusion of activated escape events with rate $1/\tau_{\rm esc}$, and (d) illustrates in addition transitions among states within $A$ (see text).
    }
    \label{fig:Illustration} 
    \end{center}
\end{figure}

Assuming that the initial state $\epsilon$ is not in the absorbing region, the qualitative features of the dynamics depends on  whether $\epsilon_\gamma$ is less or larger than the steady-state energy $\epsilon_{\rm ss}$. When $\epsilon_\gamma < \epsilon_{\rm ss}$, the evolution is ``downhill", and therefore fast, moving the mesostate energy $\epsilon$ towards larger values. 
Conversely, when $\epsilon_\gamma > \epsilon_{\rm ss}$, reaching the absorbing region requires moving ``uphill", since the dynamics on average tends to move the state away from the absorbing region and towards the steady-state $\epsilon_{\rm ss}$. Consequently, the ``uphill" relaxation times into the absorbing region are larger, and display characteristics of metastability \cite{AntonBook}, namely a fast evolution towards the region around $\epsilon_{\rm ss}$, followed by a slower, ``uphill" relaxation to the absorbing region. These features are illustrated in Fig. \ref{fig:Illustration}(a). 

We turn next to the {\em mean}  first-passage time $\tau_{\rm abs}(\gamma)$ to reach $\epsilon_\gamma$ from the steady-state $\epsilon_{\rm ss}$ in the ``uphill" regime, when $\epsilon_\gamma > \epsilon_{\rm ss}$. For the nearest-neighbour Markov chain it is given in general by \cite{vanKampen1992stochastic,AntonBook,norris1998markov,bovier2001metastability}
\begin{equation}
 \tau_{\rm abs}(\gamma) = \sum_{j = 1 + \mathcal{N}/2 }^{k_\gamma -1}\,  \frac{\mu_j}{c(j,k_\gamma)} + \frac{1}{c(\mathcal{N}/2,k_\gamma)}\, \sum_{j = 0} ^{\mathcal{N}/2} \, \mu_j,
 \label{eqn:Ehitviacap}
\end{equation}
where $\mu_j = \mu(\epsilon_j)$ is the invariant measure, and
\begin{equation}
\frac{1}{c(m,n)} = \sum_{j = m}^{n-1} \, \frac{1}{\mu_j p_{j,j+1}}.    
\label{eqn:cmn}
\end{equation}
A self-contained derivation of this result is given in the 
Supplemental Material (SM) \footnote{See Supplemental Material at {\tt <URL>}.}, which may also be consulted for additional details of the discussion below. For our present purposes, it suffices to illustrate how metastability emerges from \eqref{eqn:Ehitviacap}.
When the invariant measure $\mu_j$ is unimodal and peaked around $\epsilon_{\rm ss}$, the expressions $1/c(m,n)$ in Eq. \eqref{eqn:Ehitviacap}, due to their reciprocal dependence on $\mu_j$ in \eqref{eqn:cmn},  are dominated by the terms furthest away from $\epsilon_{\rm ss}$. This means that (i) in the regime $\epsilon_\gamma > \epsilon_{\rm ss}$, the mean first-passage time $\tau_{\rm abs}(\gamma)$ grows rapidly with $\gamma$, and more importantly (ii) that to leading order this growth depends only on the tails of $\mu$.
Thus to a good approximation the first passage time itself is exponentially distributed with mean $\tau_{\rm abs}(\gamma)$ \cite{AntonBook}.

For the Ehrenfest model the mean hitting time $\tau_{\rm abs}(\gamma)$ can be calculated directly from Eq. \eqref{eqn:Ehitviacap} and is given for $\epsilon_{\gamma} > \epsilon_{\rm ss}$ and $\mathcal{N}$ large by  \cite{bellman1951recurrence} 
\begin{equation}
 \tau_{\rm abs}(\gamma) = 2 \sqrt{\pi \mathcal{N}} \, \sqrt{ \epsilon_\gamma \left ( 1 - \epsilon_\gamma \right )}
 \, \frac{e^{2\mathcal{N}I(\epsilon_\gamma)}}{2\epsilon_\gamma - 1}, 
 \label{eqn:mk}
\end{equation}
where $I(x) = \ln 2 + x \ln x + (1 -x)\ln (1- x)$.

We turn next to the phase diagram of the Ehrenfest model. Given a fixed number $\tau$ of driving periods, 
a strain amplitude $\gamma$ and an initial state $\epsilon_{0}$, we consider whether the absorbing region $A_\gamma$ is reached within $\tau$-steps or not. The {\em phase diagram} can then be expressed in terms of the ensemble average of final energies $\epsilon_f$ reached, given $(\epsilon_{0}, \gamma)$, 
which becomes:
\begin{equation}
 \epsilon_f = \left \{
 \begin{array}{cc}
   \epsilon_0, & \epsilon_{\gamma} \le \epsilon_0, \\
   \epsilon_\gamma, & \epsilon_0 < \epsilon_\gamma \le \epsilon_{\rm ss}, \\
  \epsilon_\gamma  -  \left ( \epsilon_\gamma - \epsilon_{\rm ss} \right ) \, e^{-\tau/\tau_{\rm abs}(\gamma)}, 
   & \epsilon_\gamma > \max \left (\epsilon_0, \epsilon_{\rm ss} \right),   
 \end{array}
 \right.
  \label{eqn:genSol}
\end{equation}
The first line in Eq. \eqref{eqn:genSol} describes the case when the initial state is already in the absorbing region, while the second line corresponds to the case when the (fast) "downhill" evolution towards steady-state pushes the system into the absorbing region.
The third line describes the regime of metastability, where the system almost certainly is in the steady-state with $\epsilon \approx \epsilon_{ss}$ and the probability to reach the absorbing region within the duration $\tau$ of the driving is given as $\exp(-\tau/\tau_{\rm abs}(\gamma))$.

\begin{figure}[t!]
\begin{center}
    \includegraphics[width=\columnwidth]{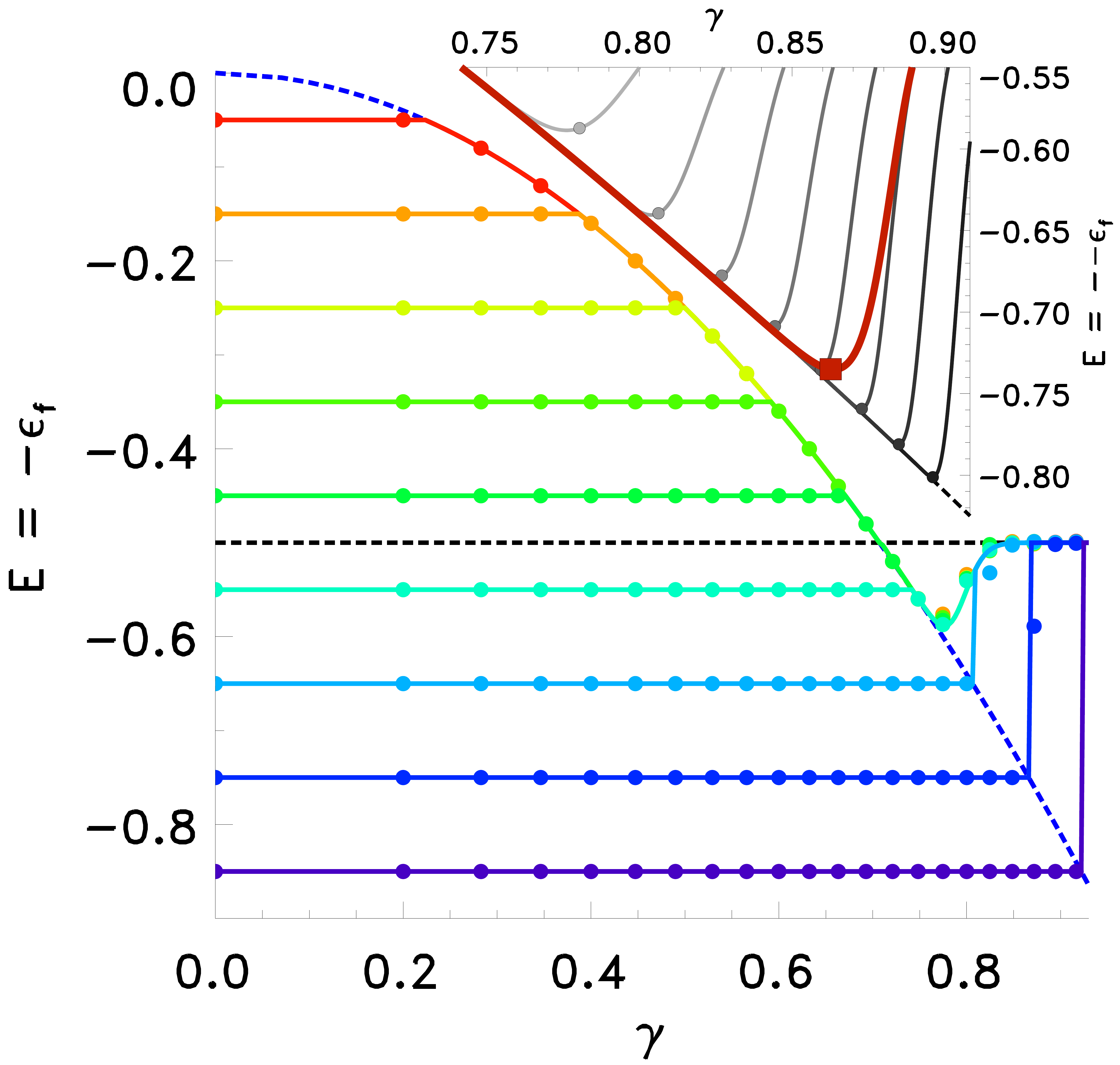} \\
    \vspace{2ex}
    \caption{The phase diagram of the Ehrenfest  model with $\mathcal{N} = 50$, and $\tau = 1000$. Each curve 
    corresponds to a sequence of final states $\epsilon_f$, averaged over $1000$ realizations,  as the driving amplitude $\gamma$ is varied, for a range of initial energy values $\epsilon_0$, each shown with symbols of a different colour. The solid lines of the same color are the theoretical predictions Eq. \eqref{eqn:genSol}. The center of the steady-state region ($E = -\epsilon_f$) is indicated by the dashed horizontal line $E = -1/2$, 
    while the blue dashed parabola $E = -\gamma^2$ marks the boundary of the absorbing region, with no dynamics. 
    The inset shows (black lines)  $\epsilon_f$, as obtained from \eqref{eqn:genSol}  with $\mathcal{N} = 50$ and driving periods $\tau = 10^3, 10^4, \ldots, 10^{10}$ (from left to right). The theoretical prediction for the minimum is indicated by the circles. The thick red line is the corresponding curve given by  \eqref{eqn:epsilonf_thermal} with finite escape rate  $\tau_{\rm esc} = 10^8$ and $\tau = 10^{10}$, 
    which determines a unique yielding point (red box), as explained in the text.}
    \label{fig:EhrenPhase}
  \end{center}
\end{figure}

Combining Eq. \eqref{eqn:mk} with Eq. \eqref{eqn:genSol}, we obtain an analytical expression describing the phase diagram $\epsilon_f(\epsilon_{0}, \gamma)$. Fig.~\ref{fig:EhrenPhase} shows the phase diagram $\epsilon_f(\epsilon_0, \gamma)$ obtained from a simulation of the Ehrenfest model with $\mathcal{N} = 50$ and $\tau = 1000$ along with the theoretical prediction (solid lines). 
Our prediction reproduces the qualitative features of the phased diagram obtained from AQS simulations of a cyclically sheared amorphous solid rather well, as can be seen by comparing Fig.~\ref{fig:EhrenPhase} with the corresponding figures 1A and 2A of \cite{bhaumik2019role}, 2a of \cite{yeh2020glass}, and 4 of \cite{liu2020oscillatory}. These features include the non-monotonic behavior of $\epsilon_f(\epsilon_0, \gamma)$ for the samples at low energies, as well as the absence of annealing up to yielding in the well-annealed samples. 
The  local minimum at $(\gamma_y,\epsilon_{\gamma_y})$ is interpreted as the onset of the yielding transition \cite{sastry2020mesoland}. 
While our results capture some aspects of the behavior above yielding, we believe that incorporation of interactions among mesoscale blocks -- not included here -- is necessary to realistically describe macroscopic behaviour in this regime.
Note that for 
$\epsilon_0 > \epsilon_{\gamma_y}$ and $\gamma \ge \gamma_y$, the numerical data points deviate from the theoretical prediction. 
The reason for this is that in the third line of Eq. \eqref{eqn:genSol}, we have assumed that the system always returns to the steady-state region before attempting to reach the absorbing region. Consequently,   
we have ignored the case when states sufficiently close to the absorbing region have an appreciable chance to reach it before relaxing to the steady state energy $\epsilon_{\rm ss}$.

The inset of  Fig.~\ref{fig:EhrenPhase} shows  $\epsilon_f$ as a function of $\gamma$ for $\tau = 10^3, 10^4, \ldots 10^{10}$
(from left to right and increasingly darker gray shade). We note that the point of upturn of $\epsilon_f$,  which we identified as the yield point, shifts to higher $\gamma$ values with increasing $\tau$ but exceedingly slowly. Indeed, from Eq.  \eqref{eqn:genSol} it is clear that at the yielding point  $\tau \sim \tau_{\rm abs}(\gamma_y) $. Using Eq.\eqref{eqn:genSol}, and employing the quadratic approximation of $I(x)$ in Eq.~\eqref{eqn:mk}, we obtain to leading order, 
\begin{equation}
    \gamma_y^2 = \frac{1}{2} \left ( 1  + \sqrt{ \frac{1 }{\mathcal{N}} \, \ln \left ( \pi^{-\frac{1}{2}} \frac{\tau}{\mathcal{N}}\right )} \right ).
      \label{eqn:yp}
\end{equation}
The theoretical prediction of the yielding point is shown in the inset of Fig.~\ref{fig:EhrenPhase} (filled circles). 
The dynamics of the mesostate models can be cast in the form of a Fokker-Planck equation in the continuum limit, as described in the SM. 

Finally, let us note that the Ehrenfest model exhibits memory formation and marginal stability \cite{coppersmith1987pulse, keimetalRMP2019}: The search for an absorbing state terminates with the first mesostate whose energy satisfies $\epsilon > \gamma^2$. Since the step size of the random walk is small, this actually happens when $\epsilon \approx \gamma^2$, so that the mesostate is barely absorbing and thus
the driving amplitude marks the boundary between absorption and diffusion, forming a memory of the strain amplitude that can be read out. 

\noindent{\it Activated Processes:} In the discussion so far, we have treated states in the absorbing region
$\epsilon > \epsilon_{\gamma}$ as being stable with respect to cycles of shear, and thus, once such a state is reached, the system remains in that state. 
This situation changes qualitatively if we consider the possibility of {\it activated} escape from such stable states.  We first consider the case of a fixed activation time $\tau_{\rm esc}$. Denoting by $P_A(t)$ and $P_S(t)$ the probabilities that the system is in an absorbing state $A$, respectively steady-state $S$, 
the evolution of this pair of probabilities follows a two-state continuous-time Markov process, 

\begin{align}
 \dot{P}_S &= - \frac{1}{\tau_{\rm abs}(\gamma)} \, P_S + \frac{1}{\tau_{\rm esc}} \, P_A, \label{eqn:P_S} \\ 
 \dot{P}_A &= - \frac{1}{\tau_{\rm esc}} \, P_A +  \frac{1}{\tau_{\rm abs}(\gamma)} \, P_S, \label{eqn:P_A}
\end{align}
with transition rates $\tau^{-1}_{\rm esc}$ and $\tau^{-1}_{\rm abs}(\gamma)$,  where $\tau_{\rm abs}(\gamma)$ is given by \eqref{eqn:mk}. Solving these with the initial condition $P_S(0) = 1$ and 
$P_A(0) = 0$, the last line of \eqref{eqn:genSol} for $\epsilon_f(\epsilon_0, \gamma)$ can now be written as 

\begin{equation}
 \epsilon_f = \mu_{S} \, \epsilon_{ss} 
  + \mu_{A}  \, 
  \left [ \epsilon_\gamma - \left ( \epsilon_\gamma - \epsilon_{ss} \right ) \, e^{-\tau/\tau_{\rm eff} } \right],
  \label{eqn:epsilonf_thermal}
\end{equation}
where $\tau_{\rm eff}^{-1 } = \tau_{\rm esc}^{-1} + \tau_{\rm abs}(\gamma)^{-1}$ and $\mu_{S} = \frac{ \tau_{\rm abs}(\gamma) } {\tau_{\rm esc} + \tau_{\rm abs}(\gamma) }$, $\mu_{A} = \frac{\tau_{\rm esc}}{\tau_{\rm esc} + \tau_{\rm abs}(\gamma) }$ are the $t \rightarrow \infty$ values of $P_{S}(t)$ and $P_{A}(t)$, respectively. The inset of 
Fig.~\ref{fig:EhrenPhase} shows the behaviour of $ \epsilon_f$ for $\tau_{\rm esc} = 10^8$ (red curve). The corresponding yield strain (red box), which becomes fixed by $\tau_{\rm esc}$, turns out to be given simply by \eqref{eqn:yp}, but with $\tau_{\rm esc}$ in place of $\tau$ (see SM). Eq.~\eqref{eqn:yp} suggests that, generically, yielding will be accompanied by a 
discontinuous change in energy, since the location of 
yield is determined by $\tau$ or $\tau_{esc}$. 

Rather than assume $\tau_{\rm esc}$ arbitrarily, it can be computed by the dynamics of the model under the assumption that an absorbing state  $\epsilon_j$ will become unstable on a time scale given by 
\begin{equation}
 \tau_j = \tau_0 \, e^{\beta \Delta E_j}.
 \label{eqn:tauj1}
\end{equation}
Here $ \Delta E_j = \frac{\kappa}{2}\, \left ( \epsilon_j - \gamma^2 \right )$ is the difference between the energy 
$E = -\epsilon_j + \frac{\kappa}{2} \epsilon_j$ of state $j$ at 
its stability limit $\gamma^{\pm} = \pm \sqrt{\epsilon_j}$
and the energy at strain $\gamma$, $E = - \epsilon_{j} +  \frac{\kappa}{2} \gamma^2$.  $ \Delta E_j$ is thus the energy barrier that must be overcome by activated processes for the state $\epsilon_j$ to become unstable and $\beta$ is the inverse temperature that determines activation rates. 
While we write the activation time in a form that corresponds to thermal activation, we do not make any specific assumptions in this work about the origin of the activation. Indeed, the idea of activation as arising from mechanical noise has been extensively studied, as also thermal noise \cite{sollich1997rheology,sollich1998,linwyart2016,parley2020}. 

Considering the steady state $S$ with measure $\mu_{S}$,  
and setting  $\epsilon_k = \gamma^2$, so that the absorbing region is formed  
by the states with $j \ge k$, the 
detailed balance conditions,  as illustrated in Fig. \ref{fig:Illustration}(d), become 
 \begin{align}
 \frac{p_{k,k-1}}{\tau_0} \mu_k - \frac{\mu_{\rm S}}{\tau_{\rm abs}(\gamma)} &= 0, \\
 \frac{p_{j+1,j}}{\tau_{j+1}} \, \mu_{j+1} - \frac{ p_{j,j+1} }{\tau_{j}} \, \mu_j &= 0, \quad j \ge k,  
\end{align}
where the first line expresses $\mu_{S}$ in terms of $\mu_k$, and for $j > k$, $\mu_j$ are given in terms of $\mu_k$ by
\begin{equation}
\mu_j = \mu_k \, \eta^{j -k} \,  \left ( \begin{array}{c} 2\mathcal{N} \\ j \end{array} \right ) \left ( \begin{array}{c} 2\mathcal{N} \\ k \end{array} \right )^{-1} , 
\end{equation}
where $\eta =   e^{ \frac{\beta \kappa}{4\mathcal{N}}}$, and $\mu_k$ is determined by normalization. Expressed in energies, $\mu_j$ nominally has a maximum at $\epsilon_{\rm max} = \frac{\beta \kappa}{16\mathcal{N}} + \epsilon_{ss}$. 
In the case $\epsilon_{\rm max} < \epsilon_\gamma$, which we do not consider further, $\tau_{\rm esc}$ is determined by the time scale $\tau_0$. 
Instead, we  assume  $\epsilon_\gamma < \epsilon_{\rm max} < 1$. Under this assumption, 
the probability $\mu_A= \sum_{j \ge k} \mu_j$ that the system is in one of the absorbing states  can be evaluated as 
\begin{equation}
  \mu_A = \mu_k \sqrt{\pi \mathcal{N}}\, e^{4 \mathcal{N} \left ( \epsilon_{\rm max} - \epsilon_\gamma \right )^2}.
 \label{eqn:muknorm2}
\end{equation}
Coarse-graining the Markov process by considering only transitions between $A$ and $S$, the dynamics reduces to \eqref{eqn:P_S} and \eqref{eqn:P_A} with the $\gamma$-dependent escape rate given by 
\begin{equation}
 \tau_{\rm esc} = \tau_0 \,  \sqrt{\pi \mathcal{N}} \, e^{4 \mathcal{N} \left ( \epsilon_{\rm max} - \epsilon_\gamma \right )^2},
\end{equation}
and providing thereby an explicit expression for $\tau_{\rm esc}$ used in Eq. \eqref{eqn:epsilonf_thermal}. Note that $\tau_{\rm esc}$ is a decreasing function of $\gamma$ whereas $\tau_{\rm abs}(\gamma)$ is an increasing function, and thus, their crossing uniquely determines the yielding point. 

We finally note that a finite and strain-dependent $\tau_{\rm esc}$ provides a mechanism by which the system may yield also well below the yielding point. This suggests the possibility that the present analysis could be extended to investigate {\it fatigue failure}, a phenomenon wherein a material may fail when subjected to repeated, cyclic loading, below the yield point. The number of cycles to such failure is known to increase exponentially with the distance away from the yielding point, a possibility supported by the results here. Extending the present analysis in that direction is the subject of future work. 

\begin{acknowledgments}
The authors would like to thank Anton Bovier, Jack Parley, Ido Regev, Ken Sekimoto, Peter Sollich, Lev Truskinovsky, and Tom Witten, for insightful discussions and careful reading of the manuscript. MM was supported by the Deutsche Forschungsgemeinschaft (DFG, German Research Foundation) under Projektnummer 398962893, the Deutsche Forschungsgemeinschaft (DFG, German Research Foundation) - Projektnummer 211504053 - SFB 1060, and by the Deutsche Forschungsgemeinschaft (DFG, German Research Foundation) under Germany’s Excellence Strategy - GZ 2047/1, Projekt-ID 390685813.
SS acknowledges support through the JC Bose Fellowship  (JBR/2020/000015) SERB, DST (India).
\end{acknowledgments}

\bibliography{AmorphNets}

\end{document}